\documentclass[twocolumn,aps,prl,superscriptaddress,nofootinbib,showpacs,floatfix]{revtex4}

\usepackage{multirow}
\usepackage{epsfig}
\usepackage{amsmath}
\usepackage{amsmath}
\usepackage{amsfonts}

\usepackage[normalem]{ulem}  
\usepackage[dvips]{color} 

\renewcommand\sout{\bgroup \color{red} \ULdepth=-.5ex \ULset}

\newcommand{\Ex}[2]{\ifmmode{#1\times10^{#2}}\else{$#1\times10^{#2}$}\fi}

\begin{document}

\title{Dirac returns: non-Abelian statistics of vortices with Dirac fermions}

\vspace{5mm}

\author{Shigehiro Yasui}
\affiliation{KEK Theory Center, Institute of Particle 
and Nuclear Studies,
High Energy Accelerator Research Organization (KEK),
1-1 Oho, Tsukuba, Ibaraki 305-0801, Japan}

\author{Kazunori Itakura} 
\affiliation{KEK Theory Center, Institute of Particle 
and Nuclear Studies,
High Energy Accelerator Research Organization (KEK),
1-1 Oho, Tsukuba, Ibaraki 305-0801, Japan}

\author{Muneto Nitta}
\affiliation{Department of Physics, and Research and Education Center 
for Natural Sciences, Keio University, 4-1-1 Hiyoshi, Yokohama, 
Kanagawa 223-8521, Japan}

\date{\today}

\begin{abstract}
Topological superconductors classified as type D 
admit zero-energy Majorana fermions inside vortex cores, 
and consequently the exchange statistics of vortices becomes 
non-Abelian, giving a promising example of non-Abelian anyons.
On the other hand,  types C and DIII 
admit zero-energy Dirac fermions inside vortex cores.
It has been long believed that an essential condition for the 
realization of non-Abelian statistics is non-locality of Dirac fermions 
made of two Majorana fermions trapped inside two well-separated vortices 
as in the case of type D. Contrary to this conventional wisdom, however, 
we show that vortices with local Dirac fermions also obey non-Abelian 
statistics. 

\end{abstract}


\maketitle

\setcounter{page}{1}
\setcounter{footnote}{0}
\renewcommand{\thefootnote}{\arabic{footnote}}

One of remarkable achievements in condensed matter physics 
in these years is the classification of topological insulators and 
superconductors \cite{SchnyderRFL:08,Kitaev:08}.
In particular, among them, the type D superconductors exhibit a unique 
property that zero-energy Majorana fermions 
\cite{Majorana:1937vz,Wilczek:2009} 
are trapped inside a vortex core \cite{Volovik:1999} 
or appear at the edge of superconductors. 
Examples of materials for type D are given by 
chiral $p$-wave superconductors/superfluids 
such as Sr$_2$RuO$_4$ and the A-phase of $^3$He.
One of important aspects of the zero-energy Majorana fermions 
is that exchange statistics of vortices 
is non-Abelian  \cite{Read:1999fn,Ivanov:2001} 
so that a set of them can be regarded as non-Abelian anyons. 
They are expected to be quite useful 
for a fault-tolerant quantum computing \cite{Kitaev:2006}. 
The idea for topological quantum computation utilizing
non-Abelian anyons on a topological state of matter
has been developed \cite{Nayak:2008zza}.   
This is why it is one of hot topics to 
 experimentally realize Majorana fermions.

In the first theoretical discovery of non-Abelian statistics of 
vortices with Majorana fermions (we call them {\it the Majorana vortices}) 
\cite{Ivanov:2001}, the key ingredient
was an unusual way to define Dirac fermions that are necessary 
for the construction  
of a Hilbert space of many vortices. With a single Majorana fermion at
each vortex core, one has to compose a Dirac fermion from two Majorana 
fermions located at spatially well-separated vortices. 
Since then, it has been long believed that the essential property 
that leads to 
non-Abelian statistics is the non-locality of Dirac fermions. 
This is one of the reasons why other types of topological 
insulators/superconductors have not attracted more attentions 
than the type D insulators/superconductors 
as candidates of non-Abelian anyons.

On the other hand, the type C and DIII insulators/superconductors 
admit Dirac fermions inside vortex cores \cite{Roy:2010}. 
We call such vortices {\it the Dirac vortices}.
There are several systems allowing Dirac fermions 
such as cores of integer (singular) vortices 
in $^3$He A-phase \cite{Kawakami:2010},  
and dislocation lines in topological insulators \cite{Imura:2011}.
Dirac fermions already exist locally in these cases,
and therefore they have been believed for a long time 
to admit no non-Abelian statistics.
Contrary to conventional wisdom, however, 
we show in the present paper that they 
actually do give non-Abelian statistics, 
and consequently it implies that 
Majorana fermions are not necessary anymore 
for constructing non-Abelian statistics. 

We treat a set of $n$ vortices, each of which allows for a single 
zero-energy Dirac fermion at its core, and consider an exchange 
of two vortices in such a system. 
An operation $T_{k}$ is defined as the anti-clockwise 
exchange of the $k$-th and $(k+1)$-th vortices.
Then, $\{ T_{k}\vert k=1,\cdots,n-1\}$ 
are in general the generators of the braid group.
They satisfy the braid relations (i) $T_{k}T_{\ell}T_{k}=T_{\ell}T_{k}T_{\ell}$ for $|k-\ell|=1$ and (ii) $T_{k}T_{\ell}=T_{\ell}T_{k}$ for $|k-\ell|>1$. 
We assume that exchange of two vortices is performed as 
an adiabatic process, and that quantum states of vortices 
are essentially determined by the zero-energy modes. Let us define 
an (annihilation) operator $\hat \psi_k$ 
for the zero-mode Dirac fermion at the $k$-th vortex. It satisfies
the following algebra: $\{ \hat\psi_k, \hat\psi_\ell^\dag  \}=\delta_{k\ell},\ 
\{\hat\psi_k, \hat\psi_\ell \}=\{\hat\psi_k^\dag, \hat\psi_\ell^\dag \}=0$.
Then, the exchange of two vortices is expressed in terms of the change 
of operators $\hat \psi_k$ under the operation $T_k$.
Note that the wave function of the zero modes obtains an additional 
phase, a minus sign, when the zero-mode fermion turns around a vortex,
implying that the zero-mode wave function is a double-valued function.
In order to regard this wave function as a single-valued function, 
we introduce a cut so that the zero-mode wave function changes its 
sign when it crosses the cut. Under the operation $T_{k}$, we find that 
the $k$-th vortex crosses the cut of the $(k+1)$-th vortex.
The operation $T_{k}$ acts on the Dirac zero-mode operators 
$\hat{\psi}_{k}$ and $\hat{\psi}_{k+1}$ 
for the $k$-th and $(k+1)$-th vortices, respectively, as
\begin{eqnarray}
T_k : \left\{ 
\begin{array}{l}
 \hat{\psi}_{k}\quad \rightarrow\, \hat{\psi}_{k+1} \\
 \hat{\psi}_{k+1} \rightarrow -\hat{\psi}_{k}  
\end{array}
\right. ,
\label{eq:exchange_Dirac}
\end{eqnarray}
with keeping the others $\hat{\psi}_{\ell}$ ($\ell\neq k, k+1$) unchanged.\footnote{The transformation (\ref{eq:exchange_Dirac}) is consistent with the 
Bogoliubov-de Gennes equation which is a microscopic equation for 
the fermions in the presence of vortices.}
We confirm that the above transformation satisfies 
the braid relations (i) and (ii). We should emphasize that, 
because of the minus sign in Eq.~(\ref{eq:exchange_Dirac}), 
we obtain the relation $T_{k}^{4}=1$, 
i.e.,  {\it four} successive exchanges are equivalent to 
the identity.\footnote{
Two successive exchanges yield $T_k^2=-1$.} 
This may be contrasted with the exchange of two conventional particles, 
in which {\it two} successive exchanges are equivalent 
to the identity.\footnote{
If there is no minus sign in Eq.~(\ref{eq:exchange_Dirac}), 
we obtain $T_{k}^{2}=1$ which implies 
the Bose-Einstein/Fermi-Dirac statistics or para-statistics.
Thus, the existence of the minus sign in Eq.~(\ref{eq:exchange_Dirac}) 
is essential. \label{footnote:Bose-Fermi}
}
This fact by itself suggests that the operation $T_{k}$ induces a 
non-trivial statistics, as discussed below.

Now, let us find the representation of the operation $T_{k}$ 
following the procedure discussed in Ref.~\cite{Ivanov:2001}.
First of all, we note that $T_k$ defined in 
Eq.~(\ref{eq:exchange_Dirac}) can be represented with respect to the Dirac fermion operator $\hat\psi_\ell$. 
The corresponding operator $\hat{\tau}_{k}$ 
reads
\begin{eqnarray}
\hat{\tau}_{k} &=& 1+ \hat{\psi}^{}_{k+1} \hat{\psi}_{k}^{\dag} 
                    + \hat{\psi}_{k+1}^{\dag} \hat{\psi}^{}_{k} \nonumber \\ 
&& - \hat{\psi}_{k+1}^{\dag} \hat{\psi}^{}_{k+1} 
- \hat{\psi}_{k}^{\dag} \hat{\psi}^{}_{k}  + 2 \hat{\psi}_{k+1}^{\dag} \hat{\psi}^{}_{k+1}\hat{\psi}_{k}^{\dag} \hat{\psi}^{}_{k} .
\label{eq:exchange_Dirac_rep}
\end{eqnarray}
We confirm that $\hat{\tau}_{k}\hat{\psi}_{\ell}\hat{\tau}_{k}^{-1}$ ($\ell=1, \dots, n$) reproduces the transformation (\ref{eq:exchange_Dirac}). 
We also note that $\hat{\tau}_{k}$ satisfies the braid relations: 
(i) $\hat{\tau}_{k}\hat{\tau}_{\ell}\hat{\tau}_{k}=\hat{\tau}_{\ell}\hat{\tau}_{k}\hat{\tau}_{\ell}$ for $|k-\ell|=1$ and 
(ii) $\hat{\tau}_{k}\hat{\tau}_{\ell}=\hat{\tau}_{\ell}\hat{\tau}_{k}$ for $|k-\ell|>1$. Moreover, we have $\hat{\tau}_{k}^{4}=1$. 
Therefore, $\hat{\tau}_{k}$ gives a representation of the braid group with a condition $\hat{\tau}_{k}^{4}=1$.\footnote{
Recall that, in the case of the {\it Majorana} vortices, 
four-times multiplication of an exchange operator is {\it projectively} 
equivalent to the identity \cite{Ivanov:2001}.}
In general, $\hat{\tau}_{k}$'s are non-Abelian; $\hat{\tau}_{k} \hat{\tau}_{k+1} \neq \hat{\tau}_{k+1} \hat{\tau}_{k}$,
which will be explicitly seen in the matrix representation of 
$\hat\tau_k$ in the Hilbert space.\footnote{In contrast, 
$\hat{\tau}_{k}$'s are Abelian when $\hat{\tau}_{k}$'s are just 
complex numbers $e^{i\theta_k}$ with real $\theta_k$'s ($\theta=0$ for bosons, 
$\theta=\pi$ for fermions, and $\theta \neq 0, \pi$ for anyons).}

Next, we note that, with the operators 
$\hat\psi_\ell \ (\ell=1,\cdots,n)$ for the zero-mode Dirac fermions, 
it is straightforward to construct the Hilbert space of the multiple vortex 
systems.
The Fock vacuum $|0\rangle$ is defined by $\hat{\psi}_{\ell}|0\rangle=0$ for 
all $\ell$, and one can construct the Hilbert space by multiplying 
$\hat\psi_k^\dag$'s on the vacuum. 
Since the fermion number $f$ is conserved under the transformation 
(\ref{eq:exchange_Dirac_rep}) (the fermion number operator 
$\hat{f}=\sum_{i=1}^{n} \hat{\psi}^{\dag}_{i} \hat{\psi}_{i}$ 
commutes with $\hat{\tau}_{k}$ for any $k$), 
the total Hilbert space $\mathbb{H}^{(n)}$ for $n$ vortices 
is decomposed into a direct sum of 
the sectors $\mathbb{H}^{(n,f)}$ with the fermion number $f$ 
($0 \le f \le n$); 
$\mathbb{H}^{(n)}=\bigoplus_{f=0}^{n}\mathbb{H}^{(n,f)}$. 
By successive multiplications of $\hat{\psi}_\ell^{\dag}$ on $|0\rangle$, 
we obtain states with the fermion number $f$. 
We use the following notation:
\begin{eqnarray}
&&|0\dots 0 1 \dots 1 \dots 1 0\dots 0\rangle =\hat{\psi}^{\dag}_{\ell_{1}} \dots \hat{\psi}^{\dag}_{\ell_{i}} \dots \hat{\psi}^{\dag}_{\ell_{f}} |0\rangle, \\
&&\ \check{1}\qquad    \check{\ell}_{1}\quad \, \check{\ell}_{i}\quad \, \check{\ell}_{f}\quad \ \, \check{n} \nonumber
\end{eqnarray}
where $\ell_{1} < \dots < \ell_{i} < \dots < \ell_{f}$; 1(0) in the $\ell$-th 
for the (non-)existence of the $\ell$-th Dirac fermion in the left hand side.
The $f$-fermion sector $\mathbb{H}^{(n,f)}$ is spanned by these states 
and has dimension $_{n}C_{f}=\frac{n!}{f!(n-f)!}$. 

The matrix representation of the operator $\hat{\tau}_{k}$ 
is also decomposed into $_{n}C_{f} \times {}_{n}C_{f}$ 
blocks ${\tau}_{k}^{(n,f)}$,  
corresponding to the $f$-fermion sectors $\mathbb{H}^{(n,f)}$. 
For example, for the basis with 0 and/or 1 for $k$-th and $(k+1)$-th Dirac fermions, the elements of the matrix ${\tau}_{k}^{(n,f)}$ are given by
\begin{eqnarray}
\langle \dots 0 0 \dots | \hat{\tau}_{k} | \dots 0 0 \dots \rangle &=& 1, \nonumber \\ 
\langle \dots 1 0 \dots | \hat{\tau}_{k} | \dots 0 1 \dots \rangle &=& -1, \nonumber\\ 
\langle \dots 0 1 \dots | \hat{\tau}_{k} | \dots 1 0 \dots \rangle &=& 1, \nonumber \\ 
\langle \dots 1 1 \dots | \hat{\tau}_{k} | \dots 1 1 \dots \rangle &=& 1,
\label{basic_structure}
\end{eqnarray}
and the others are 0.
Notice that the matrix representation is allowed for both {\it even} and {\it odd} number of vortices.\footnote{
It should be noticed that it makes sense 
to discuss the exchange of occupied ``1" and empty ``0" states which 
corresponds to the matrix elements, such as 
$\langle \dots 1 0 \dots | \hat{\tau}_{k} | \dots 0 1 \dots \rangle $. 
This is because we are considering the exchange of 
two vortices, and it is possible to have the situation where the vortex 
does not contain a Dirac zero-mode fermion in it.}
Then, the total matrix is given as ${\tau}_{k}^{(n)}=\bigoplus_{f=0}^{n}{\tau}_{k}^{(n,f)}$.
We will see below that this basic structure 
essentially determines the non-Abelian statistics of the representation.

Let us see concretely how to construct the Hilbert space $\mathbb{H}^{(n)}$
and how the non-Abelian structure emerges in the matrix expression of 
$\tau_{k}^{(n,f)}$ ($0 \le f \le n$) by using 
simpler cases with $n= 2, 3, 4$ vortices.
In the case of $n=2$, we have only one operation $T_1$.
The Hilbert space $\mathbb{H}^{(2)}$ can be
decomposed into a direct sum of three sectors 
$\mathbb{H}^{(2,0)}=\{ |00\rangle \}$, $\mathbb{H}^{(2,1)}=\{ |10\rangle, |01\rangle \}$ and $\mathbb{H}^{(2,2)}=\{ |11\rangle \}$.
Then, we obtain the matrix representation of $\hat \tau_1$ as
(see Eq.~(\ref{basic_structure}))
\begin{eqnarray}
\tau_{1}^{(2,0)} = 1, \quad 
\tau_{1}^{(2,1)} =
\left(
\begin{array}{cc}
 0 & -1 \\
 1 & 0
\end{array}
\right), \quad 
\tau_{1}^{(2,2)} = 1.
\end{eqnarray}
Here, $\tau_{1}^{(2,0)}$ and $\tau_{1}^{(2,2)}$ 
for the empty/fully-occupied sectors $\mathbb{H}^{(2,0)}$ and 
$\mathbb{H}^{(2,2)}$ are identities, while
$\tau_{1}^{(2,1)}$ for $\mathbb{H}^{(2,1)}$ contains off-diagonal components.
Even though the matrix $\tau_{1}^{(2,1)}$ is non-diagonal, 
we have only one such matrix, and thus can simply diagonalize it to
find an Abelian statistics. Indeed, $\tau_{1}^{(2,1)}$ can be diagonalized
by a unitary transformation, and we find that it has eigenvalues $\pm i$.
Therefore the exchange statistics is an Abelian anyon statistics 
in the sector $\mathbb{H}^{(2,1)}$.

Truly non-Abelian statistics emerges when the number of Dirac 
vortices is larger than or equal to $3$. 
In the case of $n=3$, the Hilbert space $\mathbb{H}^{(3)}$ can be given as a direct sum of $\mathbb{H}^{(3,0)}=\{ |000\rangle \}$, $\mathbb{H}^{(3,1)}=\{ |100\rangle, |010\rangle, |001\rangle \}$, $\mathbb{H}^{(3,2)}=\{ |110\rangle, |011\rangle, |101\rangle \}$ and $\mathbb{H}^{(3,3)}=\{ |111\rangle \}$.
We thus have the matrix representation of $\hat \tau_k$ as
\begin{eqnarray}
\tau_{1}^{(3,0)} &=& \tau_{2}^{(3,0)} = 1, \nonumber \\ 
\tau_{1}^{(3,1)} &=&
\left(
\begin{array}{ccc}
 0 & -1 & 0 \\
 1 & 0 & 0 \\
 0 & 0 & 1
\end{array}
\right), \hspace{0.3em}
\tau_{2}^{(3,1)} =
\left(
\begin{array}{ccc}
 1 & 0 & 0 \\
 0 & 0 & -1 \\
 0 & 1 & 0
\end{array}
\right),  \nonumber \\
\tau_{1}^{(3,2)} &=&
\left(
\begin{array}{ccc}
 1 & 0 & 0 \\
 0 & 0 & 1 \\
 0 & -1 & 0
\end{array}
\right), \hspace{0.3em}
\tau_{2}^{(3,2)} =
\left(
\begin{array}{ccc}
 0 & 0 & -1 \\
 0 & 1 & 0 \\
 1 & 0 & 0
\end{array}
\right), \nonumber \\ 
\tau_{1}^{(3,3)} &=& \tau_{2}^{(3,3)} = 1.
\end{eqnarray}
The empty/fully-occupied sectors $\mathbb{H}^{(3,0)}$ and $\mathbb{H}^{(3,3)}$ 
are again trivial. 
For the sectors $\mathbb{H}^{(3,1)}$ and $\mathbb{H}^{(3,2)}$
the matrices show the basic structure (\ref{basic_structure}). 
This time, it is 
not possible to simultaneously diagonalize all the matrices. Indeed, 
we find $\tau_{1}^{(n,f)}$ and $\tau_{2}^{(n,f)}$ are non-commutative; $\tau_{1}^{(n,f)} \tau_{2}^{(n,f)} \neq \tau_{2}^{(n,f)} \tau_{1}^{(n,f)}$ 
with $n=3$ and $f=1, 2$.
Therefore the exchange statistics is non-Abelian 
in the sectors $\mathbb{H}^{(3,1)}$ and $\mathbb{H}^{(3,2)}$.

We again note that, with the Dirac fermions, 
the Hilbert space is constructed for both {\it even} and {\it odd} 
numbers of vortices.
This is in contrast with the vortices with Majorana fermions \cite{Ivanov:2001}, 
in which case the Hilbert space can be constructed from only 
{\it even} numbers of vortices because a single Dirac fermion 
is constructed from two Majorana fermions belonging to different vortices.

Finally, let us also show the results for the case of $n=4$. 
We will see the basic structure (\ref{basic_structure}) and thus 
non-Abelian statistics again 
when the fermion number $f=1, 2, 3$. 
The Hilbert space $\mathbb{H}^{(4)}$ can be given as a direct 
sum of five sectors: $\mathbb{H}^{(4,0)}=\{ |0000\rangle \},$
$\mathbb{H}^{(4,1)}$
$=\{ |1000\rangle, 
|0100\rangle, |0010\rangle, |0001\rangle \}$, 
$\mathbb{H}^{(4,2)}=\{ |1100\rangle, $ 
$|1010\rangle, |1001\rangle, |0110\rangle, |0101\rangle, |0011\rangle \}$, 
$\mathbb{H}^{(4,3)}=\{ |1110\rangle, $ 
$|1101\rangle, |1011\rangle, |0111\rangle \}$, and 
$\mathbb{H}^{(4,4)}=\{ |1111\rangle \}$.
The matrix representation of $\hat \tau_k$ is:
\begin{eqnarray}
\hspace*{-2em}\tau_{1}^{(4,0)} &=& \tau_{2}^{(4,0)} = \tau_{3}^{(4,0)} = 1, \nonumber \\
\tau_{1}^{(4,1)} &=&
\left(
\begin{array}{cccc}
 0 & -1 & 0 & 0 \\
 1 & 0 & 0 & 0 \\
 0 & 0 & 1 & 0 \\
 0 & 0 & 0 & 1
\end{array}
\right), \ 
\tau_{2}^{(4,1)} =
\left(
\begin{array}{cccc}
 1 & 0 & 0 & 0 \\
 0 & 0 & -1 & 0 \\
 0 & 1 & 0 & 0 \\
 0 & 0 & 0 & 1
\end{array}
\right), \nonumber \\
\tau_{3}^{(4,1)} &=&
\left(
\begin{array}{cccc}
 1 & 0 & 0 & 0 \\
 0 & 1 & 0 & 0 \\
 0 & 0 & 0 & -1 \\
 0 & 0 & 1 & 0
\end{array}
\right),  
\end{eqnarray}
in the sectors $\mathbb{H}^{(4,0)}$ and $\mathbb{H}^{(4,1)}$, 
\begin{eqnarray}
\tau_{1}^{(4,2)} &=&
\left(
\begin{array}{cccccc}
 1 & 0 & 0 & 0 & 0 & 0 \\
 0 & 0 & 0 & -1 & 0 & 0 \\
 0 & 0 & 0 & 0 & -1 & 0 \\
 0 & 1 & 0 & 0 & 0 & 0 \\
 0 & 0 & 1 & 0 & 0 & 0 \\
 0 & 0 & 0 & 0 & 0 & 1
\end{array}
\right), \nonumber \\
\tau_{2}^{(4,2)} &=& 
\left(
\begin{array}{cccccc}
 0 & -1 & 0 & 0 & 0 & 0 \\
 1 & 0 & 0 & 0 & 0 & 0 \\
 0 & 0 & 1 & 0 & 0 & 0 \\
 0 & 0 & 0 & 1 & 0 & 0 \\
 0 & 0 & 0 & 0 & 0 & -1 \\
 0 & 0 & 0 & 0 & 1 & 0
\end{array}
\right),  \nonumber \\
\tau_{3}^{(4,2)} &=&
\left(
\begin{array}{cccccc}
 1 & 0 & 0 & 0 & 0 & 0 \\
 0 & 0 & -1 & 0 & 0 & 0 \\
 0 & 1 & 0 & 0 & 0 & 0 \\
 0 & 0 & 0 & 0 & -1 & 0 \\
 0 & 0 & 0 & 1 & 0 & 0 \\
 0 & 0 & 0 & 0 & 0 & 1
\end{array}
\right), 
\end{eqnarray}
in the sector $\mathbb{H}^{(4,2)}$, and lastly
\begin{eqnarray}
\tau_{1}^{(4,3)} &=&
\left(
\begin{array}{cccc}
 1 & 0 & 0 & 0 \\
 0 & 1 & 0 & 0 \\
 0 & 0 & 0 & -1 \\
 0 & 0 & 1 & 0
\end{array}
\right), \ 
\tau_{2}^{(4,3)} =
\left(
\begin{array}{cccc}
 1 & 0 & 0 & 0 \\
 0 & 0 & -1 & 0 \\
 0 & 1 & 0 & 0 \\
 0 & 0 & 0 & 1
\end{array}
\right),\nonumber \\
\tau_{3}^{(4,3)} &=&
\left(
\begin{array}{cccc}
 0 & -1 & 0 & 0 \\
 1 & 0 & 0 & 0 \\
 0 & 0 & 1 & 0 \\
 0 & 0 & 0 & 1
\end{array}
\right),  \nonumber \\
\tau_{1}^{(4,4)} &=& \tau_{2}^{(4,4)} = \tau_{3}^{(4,4)} = 1, 
\end{eqnarray}
in the sectors $\mathbb{H}^{(4,3)}$ and $\mathbb{H}^{(4,4)}$.
As in the previous cases, 
the empty/fully-occupied sectors $\mathbb{H}^{(4,0)}$ and 
$\mathbb{H}^{(4,4)}$ are trivial.
For the sectors $\mathbb{H}^{(4,1)}$, $\mathbb{H}^{(4, 2)}$ and $\mathbb{H}^{(4, 3)}$, we find $\tau_{1}^{(n,f)}$, $\tau_{2}^{(n,f)}$ and $\tau_{3}^{(n,f)}$ non-commutative; $\tau_{1}^{(n,f)} \tau_{2}^{(n,f)} \neq \tau_{2}^{(n,f)} \tau_{1}^{(n,f)}$, $\tau_{2}^{(n,f)} \tau_{3}^{(n,f)} \neq \tau_{3}^{(n,f)} \tau_{2}^{(n,f)}$ with $n=4$ and $f=1, 2, 3$.
Then the exchange statistics is again non-Abelian in 
the sectors $\mathbb{H}^{(4,1)}$, $\mathbb{H}^{(4,2)}$ and $\mathbb{H}^{(4,3)}$. In this way, we can construct the matrix representation of 
the exchange operators $\hat \tau_k$
for arbitrary number of Dirac vortices.

Our procedure to find the matrix representation for the exchange of two 
Dirac vortices is very similar to the one adopted by Ivanov \cite{Ivanov:2001} 
for the exchange of two Majorana vortices. However, let us clarify that 
these two are quite different in several aspects and emphasize that the 
complexity appearing in the manipulation of the Majorana vortices is not 
an essential requisite to realize the non-Abelian statistics. 
We here point out two aspects: locality of the Dirac fermions 
and conservation of the Dirac fermion number under the exchange 
of two vortices. While these two hold for the Dirac vortices as 
explicitly demonstrated in the present paper, they are violated 
in the Majorana vortices.

Concerning the locality of the Dirac fermions, we recall that 
the Dirac fermions employed to construct the Hilbert space for the Majorana 
vortices are {\it nonlocal}. More precisely, the Dirac fermion operator 
$\hat{\psi}^{{\mathrm M}}_{k}=\frac{1}{2}(\hat{\gamma}_{2k} 
+i\hat{\gamma}_{2k+1})$ is made of two Majorana operators $\hat{\gamma}_{2k}$ 
and $\hat{\gamma}_{2k+1}$ in spatially separated $2k$-th and $(2k+1)$-th 
vortices \cite{Ivanov:2001}. Namely, $2n$ Majorana vortices generate 
$n$ nonlocal Dirac fermions. In contrast, since we started with the Dirac 
vortices themselves, they are by definition
{\it local}, i.e., a Dirac fermion is located at the core of a single vortex. 

In the Majorana vortices, non-Abelian statistics appears when one 
exchanges two `unpaired' Majorana vortices, each of which is used to 
define different Dirac fermions.
Then, the exchange of two unpaired Majorana vortices induces 
non-conservation of the Dirac fermion number \cite{Ivanov:2001}.
For example, exchange in the zero Dirac fermion sector generates 
a state with two Dirac fermions. Mathematically, one finds that 
the fermion number operator 
$\hat{f}^{\mathrm{M}} = \sum_{k=1}^{n} 
\hat{\psi}^{\mathrm{M}\dag}_{k} \hat{\psi}^{{\mathrm M}}_{k}$ 
which counts the number of Dirac fermions does not commute with 
the exchange operator $\hat{\tau}^{{\mathrm M}}_{2k+1} = \frac{1}{\sqrt{2}} 
\left( 1 + \hat{\gamma}_{2k+2}\hat{\gamma}_{2k+1} \right)$ corresponding to the exchange of the unpaired Majorana vortices. 
In contrast, as mentioned before, the fermion number in the Dirac vortices 
is conserved under the exchange of two vortices.

We find interestingly enough that the Dirac operator in 
Eq.~(\ref{eq:exchange_Dirac_rep}) can be decomposed into two Majorana 
operators $\hat{\eta}_{k}^{1}$ and $\hat{\eta}_{k}^{2}$ 
($\hat{\eta}_{k}^{i\dag}=\hat{\eta}_{k}^{i}$) as 
$
\hat{\psi}_{k} = \frac{1}{2} \left( \hat{\eta}^{1}_{k} 
+ i \hat{\eta}^{2}_{k} \right),
$
up to an overall phase. 
(Notice that this expression is consistent with the exchange 
rule (\ref{eq:exchange_Dirac}), because each 
Majorana operator $\hat{\eta}^{i}_{k}$ changes the sign like 
Eq.~(\ref{eq:exchange_Dirac}) under the exchange of vortices.)
Then, the operator (\ref{eq:exchange_Dirac_rep})
can be rewritten as 
$
\hat{\tau}_{k} = \prod_{i=1,2} \frac{1}{\sqrt{2}} 
\left( 1 + \hat{\eta}^{i}_{k+1}\hat{\eta}^{i}_{k} \right).
$
This suggests that if we start with {\it two} Majorana 
fermions localized in a single vortex, then the Hilbert space of 
the Majorana vortices can be spanned by {\it local} Dirac fermions, 
and the fermion number is conserved under the exchange of vortices. 

However, if we have {\it three}, or in general, {\it odd numbers of} 
Majorana fermions in a {\it single} vortex, we cannot
construct Dirac fermions only by using the Majorana fermions in a single 
vortex, and thus necessarily encounter non-locality of Dirac 
fermions and non-conservation of fermion number. In fact, 
this was explicitly demonstrated when the number of Majorana fermions 
is {\it three} and they have  
$\mathrm{SO(3)}$ symmetry \cite{Yasui:2010yh}. The resulting non-Abelian 
statistics was found to be given by a tensor product of the matrices 
in a single Majorana case and a Coxeter group. 
An example of physical systems to realize it is a color superconductor 
\cite{Yasui:2010yw}.

Hence, we observe that non-Abelian statistics can be realized 
by using either even or odd number of Majorana fermions. When the 
fermion number is even (odd), one uses local (nonlocal) Dirac fermions,
and the fermion number operator is (is not) conserved under the exchange of 
two vortices. Therefore, we emphasize that both the non-locality 
of the Dirac fermion and the non-conservation of the fermion number are 
not essential requisite for the non-Abelian statistics, 
but the minus sign in the exchange of vortices as in 
Eq.~(\ref{eq:exchange_Dirac}) is rather important.

\begin{figure}
\includegraphics[width=1\linewidth,keepaspectratio]{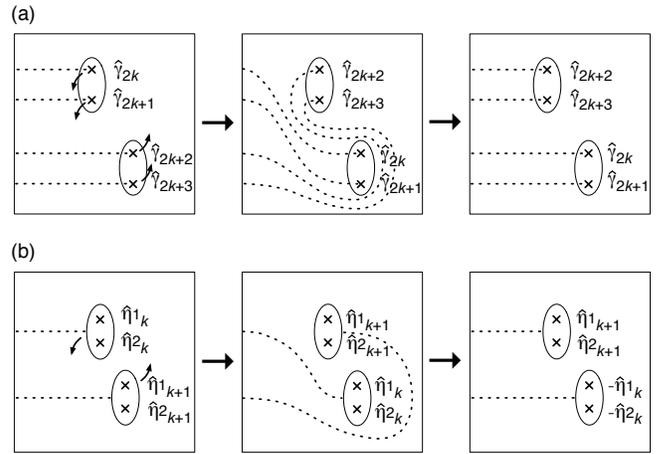}
\caption{
 The exchanges of (a) two sets of two Majorana vortices and 
(b) two Dirac vortices, 
which yield the statistics given by
Eq.~(\ref{eq:two-Majorana}) and
Eq.~(\ref{eq:exchange_Dirac}), respectively. 
They are not equivalent; the latter is non-trivial statistics 
studied in this paper, while the former is trivial statistics.
\label{fig:exchange}
} 
\end{figure}

One interesting question is if two Majorana vortices 
behave similarly to one Dirac vortex, and moreover if exchange of 
two sets of two Majorana vortices show non-Abelian statistics.
However, the exchange of two Dirac vortices 
is not equivalent to exchange of two sets of two Majorana vortices, 
see Fig.~\ref{fig:exchange}.
Consider two Dirac fermion operators in a system of (even number of) 
the Majorana vortices:
$\hat{\psi}^{{\mathrm M}}_{k}=\frac{1}{2}(\hat{\gamma}_{2k}+i\hat{\gamma}_{2k+1})$ and 
$\hat{\psi}^{{\mathrm M}}_{k+1}=\frac{1}{2}(\hat{\gamma}_{2k+2}+i\hat{\gamma}_{2k+3})$,
where the first one is made of the $2k$-th and ($2k+1$)-th Majorana vortices, 
and the second one the ($2k+2$)-th and ($2k+3$)-th Majorana vortices. 
Using the basic transformation rule for the Majorana fermions, 
one can easily find the exchange transformation of these two Dirac fermions: 
\begin{equation}
 \hat{\psi}^{{\mathrm M}}_{k} \rightarrow \hat{\psi}^{{\mathrm M}}_{k+1}, 
 \quad
 \hat{\psi}^{{\mathrm M}}_{k+1} \rightarrow \hat{\psi}^{{\mathrm M}}_{k}.
 \label{eq:two-Majorana}
\end{equation}
Thus, there is no minus sign, 
unlike Eq.~(\ref{eq:exchange_Dirac}). 
See Fig.~\ref{fig:exchange}-(a).
The matrices for exchange operators $\hat \tau_k$ in this case 
can be obtained by ignoring the minus signs in our matrices. 
It satisfies $\hat \tau_k^2=1$ giving at most parafermion statistics, 
see footnote \ref{footnote:Bose-Fermi}.

Finally we make a comment on supersymmetric gauge theories 
as a possibility of Dirac vortices. 
Bogomol'nyi-Prasad-Sommerfield vortices spontaneously break 
and preserve a half of supersymmetry in supersymmetric gauge theories.
Consequently Nambu-Goldstone fermions (Goldstino) 
for spontaneously broken supersymmetry is trapped in the core of the vortex 
\cite{Davis:1997bs}. 
When the theory posseses ${\cal N}=2$ supersymmetry (eight supercharges)
${\cal N}=1$ supersymmetry (four supercharges) 
is preserved. 
Therefore the fermion trapped in the vortex is a Dirac fermion 
because it belongs to a chiral multiplet. 
Therefore vortices in supersymmetric theories may give 
non-Abelian statistics studied in this paper. 
It remains as a future problem to study this case.

In summary, we have explicitly constructed 
the exchange statistics of vortices with Dirac fermions, 
which is relevant to the type C and DIII topological 
insulators/superconductors and some other systems, 
and have found that it is non-Abelian, contrary to 
conventional wisdom. It is expected that the existence of 
local Dirac fermions 
will provide us an interesting approach 
to observe the non-Abelian statistics 
in laboratories, helping us to utilize 
quantum computations.

\bigskip
{\bf Acknowledgements}

S.~Y. is supported by a Grant-in-Aid for
Scientific Research on Priority Areas gElucidation of New
Hadrons with a Variety of Flavors (E01: 21105006)h.
The work of M.~N. is supported in part by 
Grant-in Aid for Scientific Research (No. 23740226) 
and by the ``Topological Quantum Phenomena'' 
Grant-in Aid for Scientific Research 
on Innovative Areas (No. 23103515)  
from the Ministry of Education, Culture, Sports, Science and Technology 
(MEXT) of Japan. 


\end{document}